\documentclass[mathleft
]{an}
\usepackage{graphicx}
\usepackage{times}
\begin{document}

\Pagespan{789}{}
\Yearpublication{2006}%
\Yearsubmission{2005}%
\Month{11}%
\Volume{999}%
\Issue{88}%

\title{Variability of young stellar objects: \\
        Accretion, disks, outflows and magnetic activity}

\author{B. Stelzer\thanks{Corresponding author:
  \email{stelzer@astropa.inaf.it}\newline}
}
\titlerunning{Variability of young stellar objects}
\authorrunning{B. Stelzer}
\institute{
INAF - Osservatorio Astronomico di Palermo, 
Piazza del Parlamento 1,
90134 Palermo, Italy}

\received{10 Mar 2015}
\accepted{XXX}
\publonline{later}


\abstract{%
Variability is a key characteristic of young stars. Two major origins may be distinguished: 
a scaled-up version of the magnetic activity seen on main-sequence stars and 
various processes related to circumstellar disks, accretion and outflows.  
  }

\maketitle

\section{Introduction}

During the first few Myrs of their evolution 
late-type stars pass through several phases that are
characterized by the amount of circumstellar material and its 
interaction with the forming star. In the earliest protostellar stages the star is
surrounded by an optically thick envelope (Class\,0/I objects). Subsequently the
envelope disperses and settles into a disk (Class\,II objects) from which the star
may accrete matter (classical T\,Tauri star). Finally all circumstellar matter is
gone and only the star is left, referred to as a Class\,III object or 
weak-line T\,Tauri star. The classification with roman numbers 0 through III is related to
the slope of the spectral energy distribution in the infrared (IR) which is sensitive to the
presence of envelopes and disks, while the different T\,Tauri types represent a
spectroscopic classification measuring accretion through the strength of optical
emission lines;  
see Feigelson \& Montmerle (1999) for details. 
The definition of a "young stellar object" (YSO) sometimes comprises only the earlier 
evolutionary phases which are dominated by their circumstellar environment, 
while I include all the pre-main sequence phases mentioned above in this review
on YSO variability. 

The variability of YSOs has a variety of origins, last but
not least due to the complex and diverse environments that these objects host. 
In analogy to the Sun and late-type main-sequence stars in general, 
YSOs display a range of 
magnetic activity phenomena. These comprise coronal radiation from the radio and the X-ray bands, 
and optical and ultraviolet (UV) 
chromospheric emission lines. Stellar activity goes back to a 
dynamo mechanism, driven by convection and rotation (Noyes et al. 1984). Therefore, 
the faster rotation of young stars is associated with scaled-up activity, e.g.
$10^3$ times higher X-ray luminosity is detected from YSOs with respect
to main-sequence stars (Stelzer \& Neuh\"auser 2001; Preibisch \& Feigelson 2005). 
Flares, representing rapid release of magnetic energy through
reconnection events, are the most prominent activity signatures 
(Sect.~\ref{subsect:activity_flares}). 
Another ubiquituous
type of variability related to magnetic activity is rotational modulation produced
by inhomgeneous surface brightness related to the presence of cool star spots
(Sect.~\ref{subsect:activity_rotmod}), while
dynamo cycles (Sect~\ref{subsect:activity_cycles}) are observed mostly on old, 
inactive stars. Next to these activity-related
variability features, the presence of a circumstellar disk and mass accretion 
from the disk can 
induce variability. The matter infall from the disk to the star occurs along magnetic
field lines (Koenigl 1991). During this process the kinematic energy is transformed
into heat, leading to the formation of shocks and hot spots upon impact on the stellar surface. 
The circumstellar environment can give rise to variability either due to intrinsic changes in
mass transfer or disk emission or due to geometric effects as accretion streams 
or structures in the disk rotate in and out of the line-of-sight 
(Sect.~\ref{subsect:circumstellar_accdisk}). 
The most extreme accretion events are the
sporadic and poorly understood bursts known as EXOr and FUOr phenomena which are likely
responsible for assembling a significant amount of the final stellar mass 
(Sect.\ref{subsect:circumstellar_fuor}).
Finally, YSOs are subject to highly dynamic processes of mass loss 
(Sect.~\ref{subsect:circumstellar_jets}). 

In the remainder of this paper I review
recent developments in variability studies of the above-mentioned origins.

\section{Magnetic activity}\label{sect:activity}

\subsection{Flares}\label{subsect:activity_flares}

Stellar flares are the result of magnetic reconnection in the upper atmosphere which 
leads to particle acceleration and plasma heating. 
Footpoints of magnetic loops light up in optical/UV emission lines
and continuum following electron bombardment of the lower atmosphere.
Intense X-ray emission goes along with such events, and
is ascribed to the filling of the coronal loops with heated plasma evaporated off the
chromospheric layers along the magnetic field lines (`chromospheric evaporation', 
Antonucci et al. 1984).

The flares observed on YSOs are orders of magnitude more energetic than those
of main-sequence solar-like stars and dMe stars (so-called "flare stars"). 
Typical YSO flare 
energies in the optical broad-bands exceed $10^{35}$\,erg (Stelzer et al. 2003), 
much larger than the threshold for what is defined a "super-flare" on a solar-type 
main-sequence star (Maehara et al. 2012). 
In the X-ray band, similarly high flare energies are observed. 
However, the frequency of large flares is small ($\sim 1$ every $10\,{\rm d}$
for YSO X-ray flares; Stelzer et al. 2007a),
and sensitivity limits prevent the observation of smaller events. 
Therefore, despite the fact that the basic picture of the flare physics has been devised 
decades ago (e.g. Cargill \& Priest 1983), up to 
our days simultaneous multi-wavelength observations of stellar flares have remained 
extremely rare both for YSOs and for main-sequence stars. 
Exceptions include the first detection of a YSO flare at millimeter
wavelengths by Bower et al. (2003), observed also at centimeter bands and with serendipitous 
simultaneous X-ray coverage. 

If continuous temporal coverage and high enough signal for spectral analysis is given 
hydrodynamical (HD) models of flaring plasma loops (Reale et al. 1997) can be adapted to the 
observed time-evolution of X-ray flares, yielding
the size of the X-ray emitting structures (Getman et al. 2008a). 
Very large loop lengths, multiples of the stellar radius, have been found for YSO flares, 
and it was suspected that these events represent magnetic loops 
that connect from the star to the disk in Class\,II objects 
(Favata et al. 2005). However, after the objects' evolutionary stage was better constrained
the evidence for X-ray emitting star-disk loop structures has weakened (Getman et al. 2008b;
Aarnio et al. 2010). 

\begin{figure}
\includegraphics[width=7.0cm]{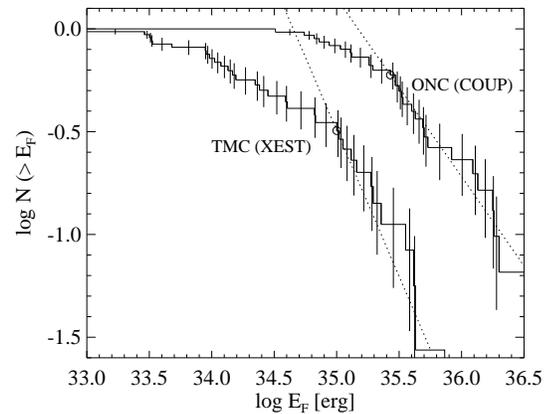}
\caption{Cumulative number distributions of X-ray flare energies 
for YSOs in the Orion Nebula Cluster (ONC) observed
during the {\em Chandra} Orion Ultradeep Project (COUP) and in the
Taurus Molecular Clouds (TMC) observed during the {\em XMM-Newton} Extended Survey
of the Taurus Molecular Clouds (XEST); from Stelzer et al. (2007a).}
\label{fig:stelzer07}
\end{figure}

The number distributions of flares are a crucial diagnostic for the importance of 
nano-flares in the coronal heating process. They have been extensively studied for the Sun
across a wide wavelength range from the UV to the hard X-ray band (see e.g. references in 
Aschwanden 2014). For YSOs the X-ray flare energies follow a power-law with 
spectral index $\approx 2$ (Stelzer et al. 2007a; 
Albacete Colombo et al. 2007; 
Caramazza et al. 2007), suggesting sufficient energy in small events to explain the
existence of the X-ray corona through nano-flaring. However, the YSO X-ray flare energy 
distributions show a turn-over for energies below $\sim 10^{35}$\,erg/s due to incompleteness
(Fig.\ref{fig:stelzer07}),
such that the nano-flare heating hypothesis relies on extrapolation over many 
orders of magnitude. Analogous to the X-ray band, 
flare energy number distributions have been studied for a small number of late-type 
main-sequence stars in the extreme UV (Audard et al. 2000) and optical bands (Hawley et al. 2014)
but equivalent studies at those wavelengths are not yet available for YSOs. 

The importance of studying X-ray flares goes beyond the scope of understanding coronal
physics. The UV and X-ray emission of YSOs is a major heating and ionization agent of 
their circumstellar disks (e.g., Gorti \& Hollenbach 2009; Ercolano et al. 2008). 
Higher-energy radiation penetrates deeper and, therefore, the effects 
on the disk crucially depend on the intensity and spectral signature of the stellar
radiation (Ercolano \& Glassgold 2013).

\subsection{Rotational modulation}\label{subsect:activity_rotmod}

Periodic optical variability of stars on time-scales of days is explained as modulation of
the surface-averaged brightness when cool and dark star spots move across the line-of-sight. 
Next to flares, these are the most frequent and well-studied stellar variability phenomenon
and a means to measure the stellar rotation rates (see e.g. Herbst et al. 2000). 
However, in Class\,II YSOs the optical emission and its variability is often dominated by
processes related to their circumstellar material from the star formation process 
(see Sect.~\ref{sect:circumstellar}). 

In the X-ray band, rotational modulation associated with structured coronae can be 
expected but is usually masked by much stronger irregular variability related to
flaring. 
A clear periodic X-ray signal persisting over several years has been observed on 
the Class\,I protostar V1647\,Ori, an EX\,Or type variable 
(see Sect.~\ref{subsect:circumstellar_fuor}), and  
the phase-folded lightcurve was modelled with 
emission from accretion hot spots (Hamaguchi et al. 2012). In fact, 
hot, bright spots representing the impact points of
accretion streams on the stellar surface may in principle produce similar rotational
signals to dark magnetic spots (see Sect~\ref{subsect:circumstellar_accdisk}). 

An nearly uninterrupted $13\,{\rm d}$-long {\em Chandra} observation -- COUP, introduced in 
Fig.~\ref{fig:stelzer07} -- 
has enabled a search for X-ray variability on rotational timescales of YSOs 
through periodogram analysis. X-ray periods were found in $\sim 10$\,\% of the studied 
sample (Flaccomio et al. 2005). 
Statistical methods examining changes of the amplitude of X-ray lightcurves throughout the 
phases of the known periodic optical variability pattern have also been used to 
establish the presence of rotational modulation in the X-ray band (Flaccomio et al. 2012). 
While amplitude peaks at optical phases $\phi \sim 0.5$ and $1.5$ 
were observed for all types of YSOs, they were most 
pronounced for Class\,II objects; see Sect.~\ref{subsect:circumstellar_accdisk} for
details.

\subsection{Dynamo cycles}\label{subsect:activity_cycles}

Dynamo cycles, similar to the Sun's $11$-yr cycle, have been observed on $\sim 60$\,\% of 
solar-type main-sequence stars through variability of the Ca\,{\sc ii}\,H\&K index 
(Baliunas et al. 1995).
For a much smaller number of late-type stars, dynamo cycles have been identified 
through photometric monitoring (e.g. Ol\'ah et al. 2009), including some young 
main-sequence stars. However, only one dynamo cycle is known on a YSO so far, the
weak-line T\,Tauri star V410\,Tau in the $\sim 2$\,Myr-old
Taurus molecular clouds. The cycle of V410\,Tau was detected in the $V$ band and has a
relatively short period of $5.4$\,yrs (Stelzer et al. 2003, Ol\'ah et al. 2009).

While broad-band photometry traces star spots, calcium emission is sensitive to 
chromospheric structures such as plages. The existing data may suggest that dynamo
cycles on younger stars are predominantly visible through spot-variations, while cycles
on more evolved stars are better probed through chromospheric tracers. 
However, strong biases -- in particular the limited amount of time invested in
the required long-term monitoring observations -- may affect our picture. 
For the same reason, unambiguous long-term periodic variability in the X-ray band 
has been identified on only four stars so far. The youngest of these
is $\approx 500$\,Myr old (Sanz-Forcada et al. 2013), clearly beyond the YSO stage. 
A possible dependence of the cycle period on age, i.e. on the overall
strength of activity, needs further investigation.

\section{Circumstellar environment}\label{sect:circumstellar}

\subsection{Accretion and disks}\label{subsect:circumstellar_accdisk}

The prime diagnostics for YSO disks and accretion are the IR and optical wavelengths. 
While disks are bright in the IR, the accretion process gives rise to optical and IR 
emission lines. 
Searching for the origin of the radiation emitted from star-disk systems 
is best pursued in the time 
domain, as different variability patterns and timescales are expected for rotational 
modulation and for intrinsic variations. 
Magneto-HD simulations predict for the stable accretion mode 
periodic broad-band lightcurves and emission line characteristics 
resulting from the changing visibility of accreting
matter in the magnetic funnel streams throughout a stellar rotation (Romanova et al. 2008;
Kurosawa et al. 2008).
These variations occur on timescales of the YSO rotation period. 
Quasi-periodic variability on time-scales of several rotation cycles is expected
in the case of structural changes of the magnetosphere as result of star-disk
reconnection (Goodson et al. 1997).
Intrinsic variations may be caused by a variety of processes including unstable
accretion through the Rayleigh-Taylor instability. The timescales for these
processes are much shorter than a rotation cycle  
and they produce stochastic variability patterns
in photometric lightcurves and emission line profiles 
(Romanova et al. 2008; Kurosawa et al. 2013).
Both photometric and spectroscopic monitoring are, therefore, powerful tools to
distinguish different variability mechanisms in YSOs. 

\begin{figure}
\includegraphics[width=7.2cm]{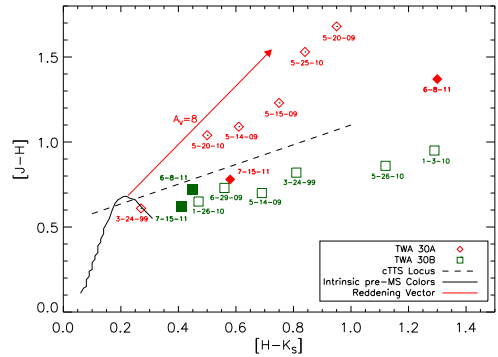}
\caption{Near-IR color-color diagram for the TWA\,30\,AB binary with data from 
Looper et al. (2010) and new observations marked with filled symbols;  
Principe et al., ApJ subm.}
\label{fig:principe}
\end{figure}

Wolk et al. (2013) showed how 
a combined analysis of near-IR magnitude and color changes and their time-scales 
helps to unveil the origin of the variability: 
Variations produced by both hot and
cool spots are periodic on timescales of the stellar rotation cycle 
but show opposite color trends; 
extinction variations are non-periodic or associated with the disk rotation and they
tend to induce dips in the lightcurve;  
variations of the mass accretion rate induce strong brightness changes. 
A case of very different near-IR variability in two stars of a young binary is 
TWA\,30\,AB (Looper et al. 2010). 
While the variations observed in TWA\,30A follow the
extinction vector suggesting obscuration effects by the disk, 
the near-IR color changes of TWA\,30B occur along
the locus populated by classical T\,Tauri stars indicating varying amount 
of dust emission from the disk. However, new observations 
challenge this clear distinction (see Fig.~\ref{fig:principe}). 

The advent of space-borne optical and IR monitoring instruments has boosted
research of photometric YSO variability associated with disks and accretion. 
Weeks or even months of uninterrupted data have become available from 
space missions such as {\em Spitzer} and {\em CoRoT} (e.g., Morales-Calderon et al. 2011; 
Cody et al. 2014). 
The morphology of irregular variations allows e.g. to distinguish stellar 
flares from accretion bursts (Stauffer et al. 2014). 
Performing such observations for a significant number of YSOs -- as can
readily be achieved with the large field-of-views of 
imaging instruments given the clustered nature of 
YSOs -- yields information on the relative frequencies of different types of 
variability patterns (Cody et al. 2014). 

Variable mass accretion, whether intrinsic or caused by geometric effects, has been cited
as a possible cause of the large observed spread in the relation between mass accretion
rate ($\dot{M}_{\rm acc}$) and stellar mass (e.g. Mohanty et al. 2005). 
However, recent results on a 
large sample of accreting YSOs indicate that the spread of $\dot{M}_{\rm acc}$
for given mass is significantly larger than what can be accounted for by variability
(Venuti et al. 2014). The accretion rates determined in this study have been obtained
from photometric measurements of UV excess emission shown by Gullbring et al. (1998)
to be correlated with $\dot{M}_{\rm acc}$. 
Alternative diagnostics for $\dot{M}_{\rm acc}$ are optical and near-IR emission lines. 
The H$\alpha$ line is usually the brightest
emission feature in YSO spectra, and therefore the most widely used probe of accretion 
despite the fact that its profile has long been recognized to comprise contributions from outflows 
(Calvet 1997). This is the likely reason why accretion rate estimates from H$\alpha$
emission yield vastly different $\dot{M}_{\rm acc}$ values than other 
prominent accretion signatures, such as higher-n Balmer lines or calcium 
(\ion{Ca}{\sc ii}\,H\&K and \ion{Ca}{\sc ii}\,IRT) and helium 
(e.g., \ion{He}\,{\sc i}$\lambda 587$, \ion{He}\,{\sc i}$\lambda 1083$) 
lines (Alcal\'a et al. 2014). 

\begin{figure}
\includegraphics[width=5.4cm,angle=270]{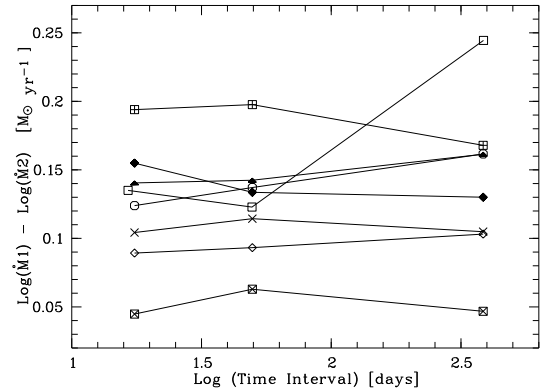}
\caption{Average spread in accretion rate for 8 Class\,II YSOs in the Cha\,I star forming
region over various timescales
showing little increase beyond the shortest timescale probed ($8...25$\,d); 
from Costigan et al. (2012).}
\label{fig:costigan12}
\end{figure}

For a given star, the individual emission lines can show different variability patterns 
(e.g. Donati et al. 2008; Stelzer et al. 2014) suggesting different formation regions. 
In a long-term, low-cadence spectroscopic monitoring program of a small sample of YSOs 
the amplitude 
of emission line variability reached a maximum after $25$\,d, indicating that the
dominant timescales for accretion variability is $\leq 2-3$ weeks (Fig.\ref{fig:costigan12};
Costigan et al. 2012). 
On a shorter observing cadence, both slow line profile variations suggestive of rotational 
modulation and rapid events possibly related to disk or magnetosphere instabilities
were observed (Costigan et al. 2014). Similarly, dedicated H$\alpha$ monitoring of the
young benchmark brown dwarf 2MASSW\,J1207334-393254 showed both smooth variations on a 
timescale corresponding
to the rotation period and fast variations implying drastic changes of the mass accretion
rate (Stelzer et al. 2007b; Scholz et al. 2005). The similarity of the variability pattern
with stellar-mass accreting YSOs underlines that substellar objects undergo a T\,Tauri phase.

Multi-wavelength monitoring combining photometric and spectroscopic observations
has the potential to connect variations related to accretion (emission lines) to 
those related to disk emission (IR brightness). Such campaigns are technically 
challenging, have therefore been limited to few objects and have so far not provided clear
results (Faesi et al. 2012) except for individual cases.
One of the best testcases for YSO variability is AA\,Tau, a classical T\,Tauri star, 
which thanks to its known, high inclination ($i \sim 75^\circ$) is ideally suited to 
demonstrate the complexity of the YSO circumstellar environment.
Among the characteristic features of AA\,Tau are 
(i) rotationally modulated veiling variations indicating the presence of hot, 
accretion spots, 
(ii) time-delays between emission line peaks and veiling variations,
(iii) photometric brightness variations interpreted as occultations by a disk warp, 
and (iv) a recent fading of the optical brightness without associated changes of its 
$\dot{M}_{\rm acc}$ and attributed to enhanced dust extinction from a density
perturbation in the disk 
(Bouvier et al. 2003; Bouvier et al. 2013). 

There are manifold interactions between the YSO X-ray emission resulting from magnetic activity
on the stellar surface and the circumstellar environment. 
First, as mentioned in Sect.~\ref{subsect:activity_rotmod}, differences have been observed 
in the X-ray variability pattern of YSOs with and without disks/accretion. 
Flaccomio et al. (2012) ascribed them to
time-variable absorption of the coronal X-ray emission by structures in the circumstellar 
material, such as disk warps or accretion streams. 
The same phenomenon may also explain the on average factor-two deficiency of the X-ray
luminosity of Class\,II objects with respect to YSOs without disks (Gregory et al. 2007).   

While the presence of a YSO disk may affect the visibility of the stellar corona as
seen above, 
an influence of X-ray emission on disk structure and accretion has been advocated 
in several theoretical works. Ke et al. (2012) predicted 
the enhanced ionization of disk grains resulting from the strong X-ray flux during
YSO coronal flares to produce changes in the height of the inner disk rim 
and associated IR variability. In a different scenario, flares in 
the star-disk magnetosphere provoke a density perturbation in the disk which gives rise
to a mass accretion event (Orlando et al. 2011). 
No convincing correlation between X-ray and IR variability in YSOs has been observed so far
(Flaherty et al. 2014), and the relation between X-ray properties and $\dot{M}_{\rm acc}$ is
unexplored to date. 
However, the typical low cadence of X-ray observations may have
prevented the detection of joint variations with optical/IR light on short timescales. 
An exception is the case of TW\,Hya on which 
Dupree et al. (2012) have identified episodes of enhanced mass accretion in soft X-ray
emission produced in the accretion shock\footnote{YSO X-ray emission at energies $\leq 0.5$\,keV may be produced in accretion 
shocks which can be heated by infalling matter up to $\sim 3-4$\,MK. 
The tendency of Class\,II
objects to show an excess of soft X-rays with respect to Class\,III objects was 
understood as evidence for accretion -- rather than the corona -- playing a major role in 
this X-ray component (G\"udel \& Telleschi 2007).}. 
The observed soft X-ray variability of TW\,Hya was contemporaneous with profile 
variations in Balmer lines indicating downflow towards the star, 
and they were followed by an increase in veiling with a few-hours delay
compatible with the expected inflow timescale. Moreover, the increase in veiling was
correlated with the harder ($>1$\,keV) coronal X-ray flux, supporting the picture of an 
"accretion-fed corona" where the stellar X-ray emission is enhanced as a result of 
heating by the accreting matter (Brickhouse et al. 2010).

\subsection{FUOr and EXOr events}\label{subsect:circumstellar_fuor}

A small number of YSOs has been observed to exhibit large brightness variations
(up to $4-5$ mag). These objects are classified according to the timescales of the
outbursts as FU\,Ors (duration of decades; Hartmann \& Kenyon 1985) or as EX\,Ors
(duration of months and recurrent; Herbig 1989). 
Both types of outbursts are ascribed to drastic accretion rate changes.  
A major part of the stellar mass may be assembled during such outbursts
but the physical mechanisms triggering the outburst are still debated;  
see Audard et al. (2014) for a recent review.

Contradictory results have been obtained from the study of YSO X-ray emission and its
variability during accretion outbursts. While the X-ray lightcurves of EX\,Ors tend to be 
correlated 
with the optical/IR flux, no clear picture has emerged yet for the characteristics and
origin of the X-ray emission, and only a handful of objects have been monitored in X-rays. 
In most of these, the temperature of the X-ray emitting gas 
was found to be too high for being generated in accretion shocks,  
but see Teets et al. (2012) for a different example. An alternative interpretation for the 
softening of the X-ray emission observed during some optical/IR outbursts is a 
change in the structure of the X-ray emitting magnetosphere caused by the accretion event 
(Audard et al. 2005).

\subsection{Jets and outflows}\label{subsect:circumstellar_jets}

Mass ejection is an essential ingredient of pre-main sequence evolution and might be 
responsible for carrying away the excess angular momentum of the accreted matter. 
The highly collimated, fast moving ($\sim 100-1000$\,km/s) inner parts are referred to
as "jets" and the wide-angle slowly moving ($\sim 10$\,km/s) molecular ejecta as "outflows";
see Frank et al. (2015) for a review. 
YSO jets evolve on timescales of few years and are, thus, easily
accessible to variability studies. Since the first proper motion measurements of 
Herbig \& Jones (1981), the bow-shocks associated with YSO jets -- so-called Herbig-Haro 
objects -- have been imaged for YSOs in a wide range of evolutionary stages 
(Class\,0 to Class\,II). In particular, narrow-band imaging with the {\em Hubble Space
Telescope} thanks to its high spatial resolution 
has enabled detailed studies of the morphology and dynamics of YSO outflows (see e.g.
Hartigan et al. 2011).  
Next to the propagation of individual knots, variability due to jet precession or orbital
motion in binaries can be tracked through optical and IR imaging by studying the 
displacements of different knots from a linear axis pointing towards the star (e.g.
Cunningham et al. 2009). These 
variations occur on secular timescales and can not be traced by multi-epoch observations.

\begin{figure}
\parbox{8.3cm}{
\parbox{4.0cm}{
\hspace*{0.2cm}
\includegraphics[width=3.6cm]{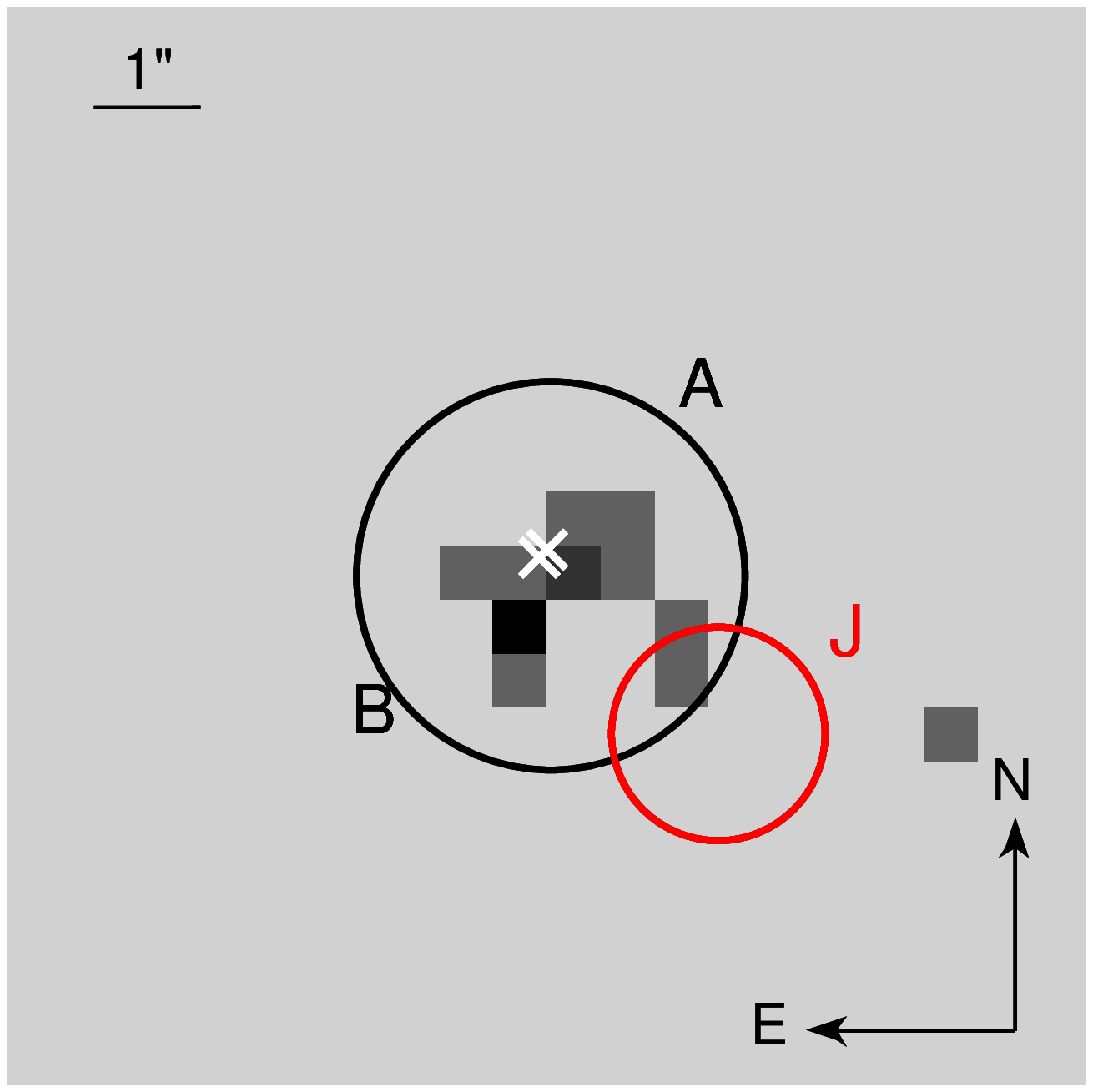}
}
\parbox{4.0cm}{
\hspace*{0.3cm}
\includegraphics[width=3.6cm]{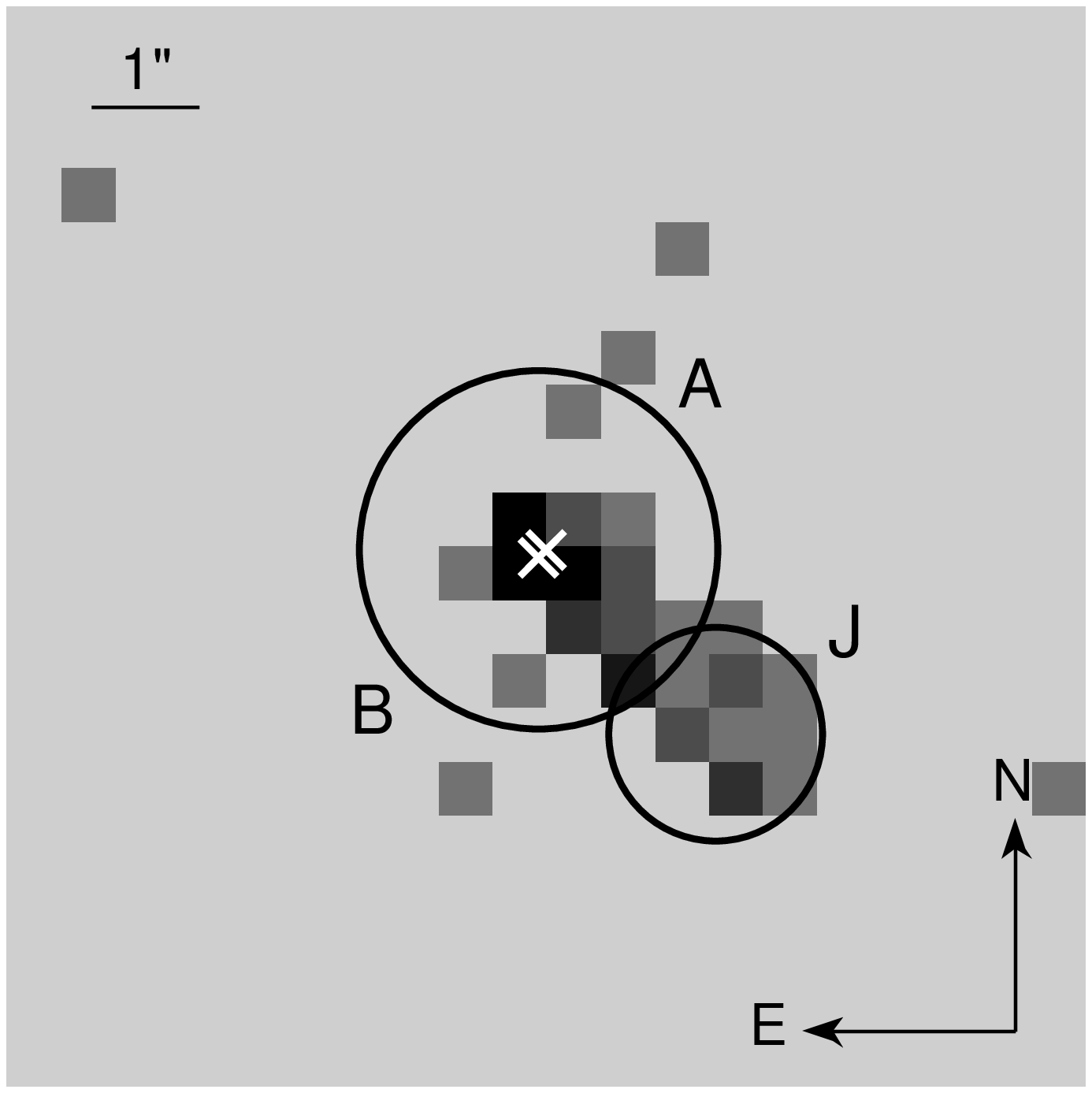}
}
}
\caption{{\em Chandra}/ACIS images for the YSO Z\,CMa in the $0.2-1$\,keV band 
 obtained in Dec 2003 (left) and in Dec 2008 (right). 
The detected X-ray sources are marked with black circles. The unresolved binary 
components of Z\,CMa are indicated with white x-shaped symbols. The position of the X-ray source detected in the 2008 
image to the SW of Z\,CMa, labeled `J',  is overlaid in the 2003 image as red 
circle. In 2008, the soft X-ray image shows elongated emission along the optical jet axis. 
(Figure adapted from Stelzer et al. 2009).}
\label{fig:stelzer09_f2}
\end{figure}

Roughly a dozen YSO jets have been detected in X-rays so far (see summary in 
Bonito et al. 2007). 
Variations of the source positions and strength in X-ray images taken a few years apart 
were observed in a few of these (Favata et al. 2006; Stelzer et al. 2009 and 
Fig.~\ref{fig:stelzer09_f2}; G\"udel et al. 2011). 
For typical shock velocities on the order of $500$\,km/s, jet proper motion can be detected
in the most nearby YSOs ($\sim 100$\,pc) in less than a decade. 
Yet it is unclear to date if the X-ray knots represent 
propagated plasma ejected and heated several years prior to the observation 
or transient emission from locally heated physical structures (Schneider et al. 2011). 
Short predicted cooling times (G\"uenther et al. 2009) favor the latter interpretation. 
On the other hand, the outward decrease of
density results in a decrease of heating efficiency such that X-ray jets are expected
to be most luminous close to the star. In fact, the X-ray jets imaged so far were found at
separations $\leq 5^{\prime\prime}$ from the central driving source, resolvable at
present only with {\em Chandra}.

The observational evidence of both stationary and moving components in X-ray jets has been 
addressed in two-dimensional HD simulations in which 
the X-ray emission from a supersonic jet interacting with its ambient medium is synthesized. 
Bonito et al. (2010) 
showed that a pulsed jet with different values for crucial jet parameters, such as 
density and ejection rate, can describe the 
X-ray luminosities and temperatures of some of the observed YSO jets but it fails to 
explain the low X-ray temperatures of the knots with the largest separation from the
star ($> 1000$\,AU). 
The stationary (non-variable) X-ray component seen in some jets can be reproduced if 
 a nozzle at the base of the jet is introduced into the model (Bonito et al. 2011). 
Currently our understanding of the high-energy component of YSO jets
is limited mainly by the large uncertainties of the observed X-ray parameters 
imposed by sensitivity constraints.


\end{document}